\documentclass[
preprint,
floatfix,
aps,
prl,
superscriptaddress,
amssymb,
]{revtex4}

\usepackage{float}
\usepackage{amssymb}
\usepackage{latexsym}
\usepackage{graphicx}
\usepackage{amsmath}
\usepackage{graphicx}
\usepackage{dcolumn}
\usepackage{amsfonts}
\usepackage{bm}
\usepackage{epsfig}

\newcommand{\be}{\begin{equation}}
\newcommand{\ee}{\end{equation}}
\newcommand{\bea}{\begin{eqnarray}}
\newcommand{\eea}{\end{eqnarray}}

\newcommand\fpk{\mbox{$f_{\mathrm{pk}}$}}
\newcommand\resxx{\mbox{Re\,$(\sigma_{\mathrm{xx}})$}}
\newcommand\smax{\mbox{$\sigma_{\mathrm{max}}$}}
\newcommand\spk{\mbox{$\sigma_{\mathrm{pk}}$}}
\def\sf{S/f_{\mathrm{pk}}}

\newcommand\na{\mbox{$\nu_{1}$}}
\newcommand\nb{\mbox{$\nu_{2}$}}

\def\to{t_{\mathrm{0}}}
\def\zo{Z_{\mathrm{0}}}
\def\lb{\ell_{\mathrm{B}}}
\def\dsas{\mbox{$\Delta_{\mathrm{SAS}}$}}

\newcommand{\rfig}[1]{Fig.\,\ref{#1}}
\newcommand{\rFig}[1]{Figure\,\ref{#1}}

\begin{document}
\title{Microwave spectroscopic studies of the bilayer electron solid states at low Landau filling in a wide quantum well
}

\author{A.\,T. Hatke}
\email[Corresponding author: ]{hatke@magnet.fsu.edu}
\affiliation{National High Magnetic Field Laboratory, Tallahassee, Florida 32310, USA}

\author{Y.\,Liu}
\affiliation{Department of Electrical Engineering, Princeton University, Princeton, New Jersey 08544, USA}

\author{L.\,W. Engel}
\affiliation{National High Magnetic Field Laboratory, Tallahassee, Florida 32310, USA}
 
\author{M. Shayegan}
\affiliation{Department of Electrical Engineering, Princeton University, Princeton, New Jersey 08544, USA}

\author{L.\,N. Pfeiffer}
\affiliation{Department of Electrical Engineering, Princeton University, Princeton, New Jersey 08544, USA}

\author{K.\,W. West}
\affiliation{Department of Electrical Engineering, Princeton University, Princeton, New Jersey 08544, USA}

\author{K.\,W. Baldwin}
\affiliation{Department of Electrical Engineering, Princeton University, Princeton, New Jersey 08544, USA}
 
\received{\today}

\maketitle

%\section{Introductory Paragraph}
\textbf{   
At the low Landau filling factor $(\nu)$ termination of the fractional quantum Hall effect (FQHE) series,  two-dimensional electron systems (2DESs) exhibit an insulating phase that is understood as a form of pinned Wigner solid \cite{lozo:1975,andrei:1988,goldman:1990,reentrant,williams:1991,kunwc,archer:2013,msreview}.   
Here we use microwave spectroscopy to probe the transition to the insulator for a wide quantum well (WQW) sample that can support single-layer or bilayer states depending on its overall carrier density, $n$.   
We find the insulator exhibits a resonance, which is characteristic of a bilayer solid.   
The resonance also reveals a pair of transitions within the solid, which are not accessible to dc transport measurements.   
As $n$ is biased deeper into the bilayer solid regime, the resonance grows in specific intensity, and the transitions within the insulator disappear.   
These behaviors are suggestive of a picture of the insulating phase as an emulsion of liquid and solid components.
}

\begin{figure}[p]
\includegraphics[width=0.48\textwidth]{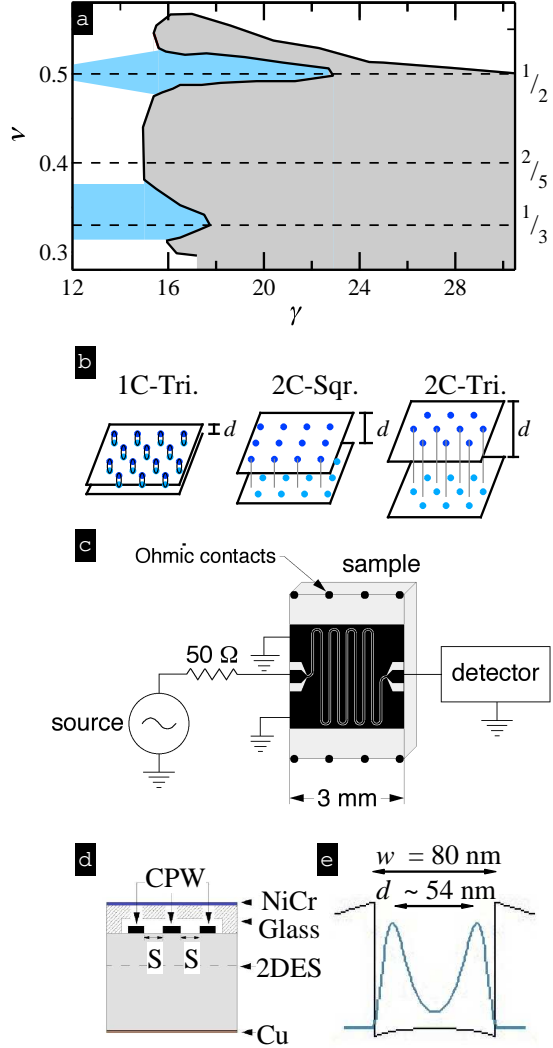}\vspace{-0.2 in}
\caption{\baselineskip 13pt   \textbf{Wide quantum well and microwave measurement.} 
\textbf{a}, Phase diagram in the $\nu$-$\gamma$ plane, where $\nu$ is the Landau filling and  $\gamma$ is the Coulomb energy ($e^{2}/4 \pi\epsilon_0\epsilon\lb$) divided by the interlayer tunneling gap, as discussed in the text.   
The insulating area is shaded grey, and the ranges of the FQHE states at $\nu=1/2$ and $\nu=1/3$ are shaded blue.
The diagram is from  the dc transport studies of Refs.\,\citep{manoharan:1996,mssemi}.   
\textbf{b},  Different possible structures for bilayer Wigner solids  as layer separation  increases for fixed $n$.  A one-component (1C) triangular structure (left), proceeds to a two-component  (2C) staggered square structure (middle), and then to a 2C  staggered triangular structure (right), adapted from Ref.\,\citep{narasimhan:1995}. 
\textbf{c}, Schematic of the microwave measurement set-up.
The source and detector are outside the cryostat at room temperature and the coplanar waveguide transmission line is patterned in metal film on top of the sample surface.   
The microwave conductivity $\sigma_{xx}$ is calculated from loss through the line as described in the Methods section. 
\textbf{d}, The microwave set-up shown in a cutaway side view.
Slots of width $s$ separate the center, driven conductor and the ground planes.  
Details of the measuring technique are given in the Methods section. 
\textbf{e}, The growth-direction electron charge distribution for a well of width $w=80\,$nm at $n=1.26\times 10^{11}$cm$^{-2}$ obtained from one-dimensional simulations.   
The growth-direction separation of charge density  peaks is $ \sim 54\,$nm.
}
\label{fig1}
\end{figure}

WQWs support bilayer as well as single-layer states, depending on the well width, $w$, and  on  the overall areal carrier density, $n$ \citep{suen:1991,suen:1992,suen:1992b,suen:1994,manoharan:1996,mssemi}. 
A measure of the tendency of the charge in a WQW to separate into two layers is $\gamma\equiv(e^{2}/4 \pi\epsilon_0\epsilon\lb)/\dsas$ \citep{suen:1994,manoharan:1996,mssemi} where $\lb=(\nu/2\pi n)^{1/2}$ is the magnetic length, $\dsas$ is the interlayer tunneling gap, and $e^{2}/4 \pi\epsilon_0\epsilon\lb$ is the Coulomb energy.   
\rFig{fig1}\,(a) shows a phase diagram in the $\nu$-$\gamma$ plane.
The interpretation of the insulator as a bilayer solid was inferred in Ref.\,\citep{manoharan:1996}  from the phase diagram  and from the response of the insulator  to asymmetric gate bias that produced mismatched   layer densities.
At small $\gamma$ the 2DES  behaves like a single layer, exhibiting a one-component (1C) Wigner solid for $\nu$ below the 1/5 FQHE, and also for a narrow reentrant range \cite{reentrant}  above it.      
At large $\gamma$ the 2DES behaves as two layers with weak interaction, and the solid occurs at $\nu\sim 2/5$, or per-layer filling $\nu_L\sim 1/5$.
At intermediate $\gamma$ the insulator includes  a $\nu$ range reentrant above the 1/2 FQHE \cite{manoharan:1996,mssemi}, which is known to be a bilayer, interlayer-correlated state, so the nearby insulator is likely a bilayer two-component (2C) solid.

Microwave spectroscopy is ideal for studies of electron solids  since these   exhibit pinning mode resonances \citep{fukuyama:1978,andrei:1988,williams:1991,doveston,chen:2004,zhw,zhwimbal,hatke:2014}, in which pieces of the solid oscillate within the potential of the residual disorder.  
This disorder also pins the solid, rendering it insulating.    
The resonance peak frequency, $\fpk$, is sensitive to  solid  properties, such as  shear modulus and the proximity of the carriers  to disorder in the host.      
A recent  study of WQW pinning modes \citep{hatke:2014} focused on the small-$\gamma$, single-layer regime in the neighborhood of $\nu=1$.
Earlier studies of bilayer pinning modes \cite{doveston,zhw,zhwimbal} used double quantum wells, for which the layers are defined by a barrier, and 1C-2C transitions were considered in Refs.\,\cite{zhw,zhwimbal}.

%%%%%%%%%%%%%%%%%%%%%%%%%
%fig 1
%%%%%%%%%%%%%%%%%%%%%%%%%

Our measurements were performed on  a GaAs/AlGaAs WQW of width $w=80\,$ nm with an as-cooled density of $n=1.1$ in units of $10^{11}\,$cm$^{-2}$ which we use throughout the paper for brevity.  
The microwave  technique  \citep{chen:2004,zhw,zhwimbal,hatke:2014} is diagrammed in Figs.\,\ref{fig1}\,(c) and (d). 
We maintained a symmetric growth-direction charge distribution about the well center   unless otherwise noted,  and calculated $\gamma(n)$ from simulations (see Methods section).     
The sample was measured in a  $60\,$mK bath.

%%%%%%%%%%%%%%%%%%%%%%%%%
%fig 2
\begin{figure}[p]
\includegraphics[width=0.48\textwidth]{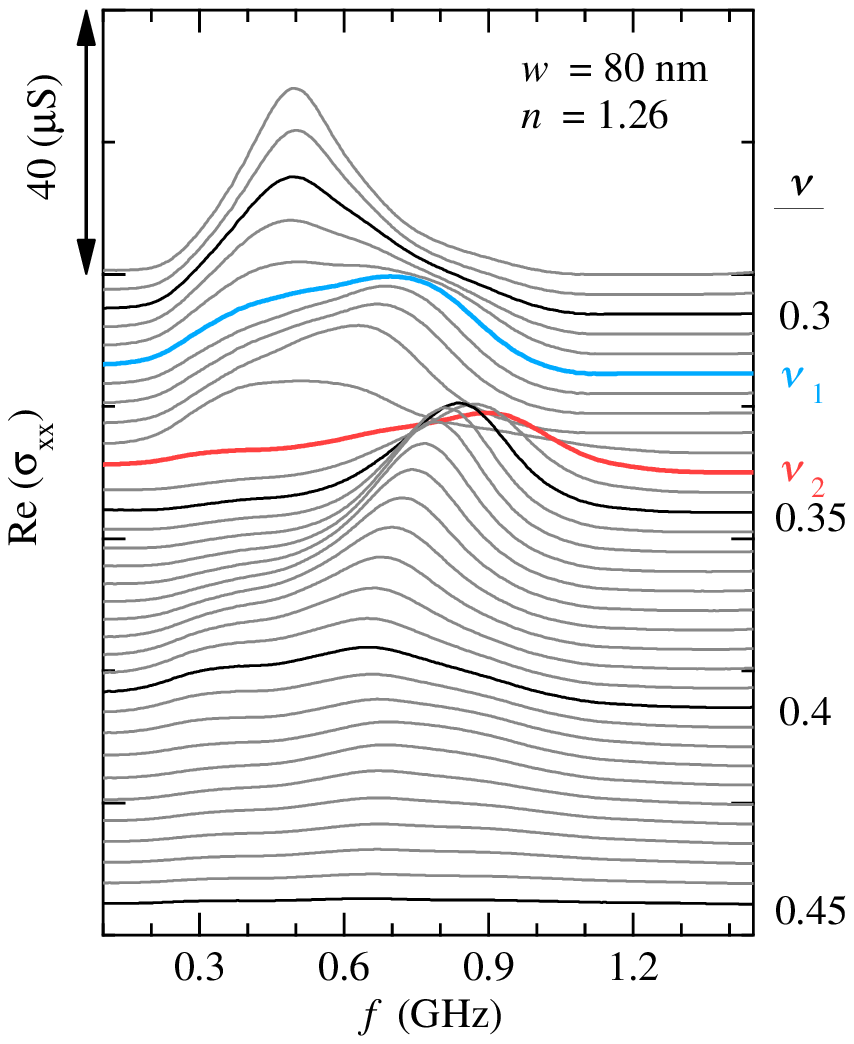}
\vspace{-0.2 in}
\caption{\textbf{Two transitions in microwave spectra.}
%MS: changed the ordering around to help with the flow
Microwave spectra, Re\,$(\sigma_{xx})$, vs frequency, $f$, at fixed $\nu$ from $0.29$ (top) to $0.45$ (bottom) with step of $0.005$ as marked along the right axis, at $n=1.26$. 
The spectra at low $\nu$ are dominated by a pinning mode resonance attributed to an electron solid.
The spectra suggest the occurrence of two transitions within the solid, one at $\nu=0.315$, the other at $\nu=0.34$.
At each transition the peak frequency, $\fpk$, exhibits an abrupt upward jump, and the amplitude of the resonance exhibits a minimum. 
The broad weak peak at $\sim 0.3\,$GHz is an  artifact due to a reflection near the sample mounting.
}
\label{wf}
\end{figure}
%%%%%%%%%%%%%%%%%%%%%%%%%

%%%%%%%%%%%%%%%%%%%%%%%%%
%fig 3
\begin{figure}[p]
\includegraphics[width=1\textwidth]{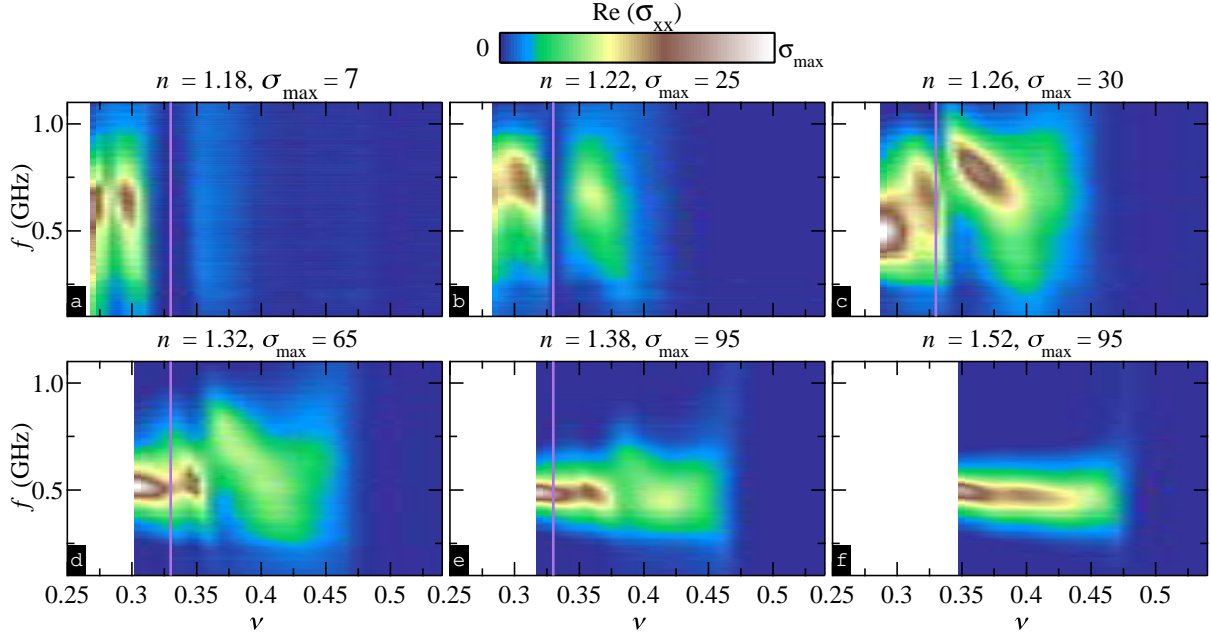}
\vspace{-0.2 in}
\caption{\textbf{Electron density ($n$) dependence of the  transition.}
Plots of the microwave spectra, Re\,$(\sigma_{xx})$, in the $(\nu,f)$-plane at various densities, $n$.
\smax\ is the largest Re\,$(\sigma_{xx})$ on the color scale.  
Vertical lines mark $\nu=1/3$ and the solid white area at low $\nu$ is outside the magnetic field measurement range. 
\textbf{a}, For $n=1.18$, the FQHE around $\nu=1/3$ appears as a region with a flat spectrum and vanishing Re\,$(\sigma_{xx})=0$.  
A resonance is seen only for  $\nu\leq 0.31$, just below the FQHE state.    
At $\nu=0.285$, the resonance peak (\spk ) in Re\,$(\sigma_{xx})$ has a sharp minimum, interpreted as a transition within the solid.  
Possibly coincidentally, this $\nu$ is within error of 2/7.     
\textbf{b}, On increasing to $n=1.22$, the resonance is reentrant around the 1/3 FQHE and a minimum in \spk\ is located at $\nu=0.300$.   
%Relative to the data of panel (a), the FQHE width in $\nu$ is reduced, because $n$ is just below that of the transition to the insulator at $   \nu=1/3$, see (c).
\textbf{c}, This data, for  $n=1.26$, is the same as   in  Fig.\,\ref{wf}.  
A resonance has replaced the FQHE at $\nu=1/3$, while for $\nu=0.315$ and $0.340$ there are \spk\ minima, which are accompanied by jumps in $\fpk$.  
\textbf{d}, \textbf{e}, Data for $n=1.32$  and $n=1.38$, respectively.   
Each plot shows two minima in \spk\ with jumps in $\fpk$.  
The jump in \fpk\ is larger for the higher-$\nu$ (\nb) minimum in \spk\ .
For $\nu$ above $\nb$ the resonance is broader with lower \spk.   
The jump in \fpk\ for the lower-$\nu$ (\na) transition at $n=1.38$ is much weaker than it is for lower $n$.   
\textbf{f}, Data for $n = 1.52$, at which both transitions have disappeared, and  $\fpk$ and the resonance amplitude vary only marginally throughout the measurement range.
}
\label{iplots}
\end{figure}
%%%%%%%%%%%%%%%%%%%%%%%%%

%%%%%%%%%%%%%%%%%%%%%%%%%
%fig 4
\begin{figure}[p]
\includegraphics[width=1\textwidth]{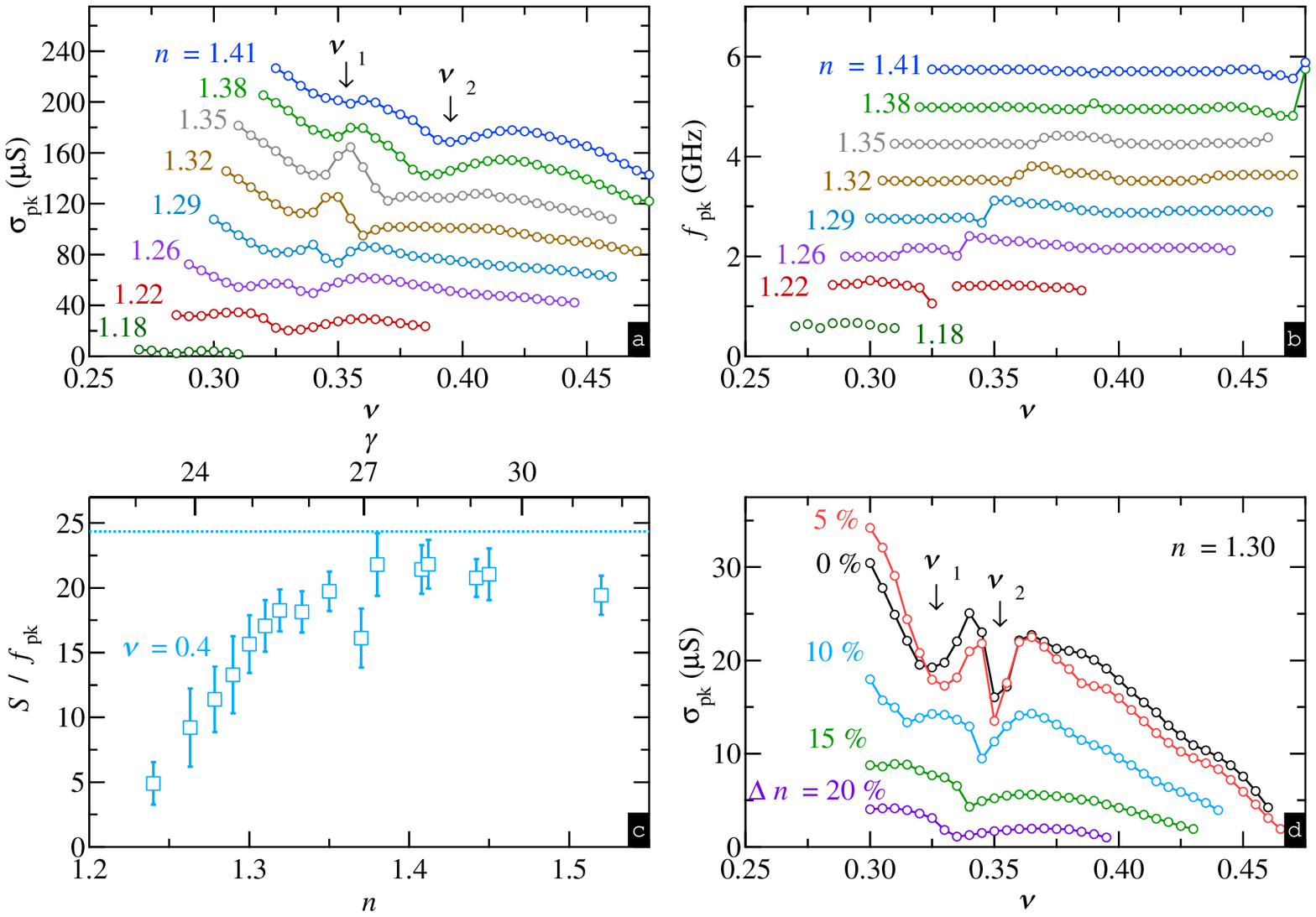}
\caption{\textbf{Electron density ($n$) and Landau filling ($\nu$) dependence of resonance parameters.}
\textbf{a}, The resonance peak conductivity,  \spk,  vs $\nu$ for many $n$ successively offset by $20\,\mu$S.
\textbf{b}, The peak frequency $\fpk$ vs $\nu$ for the same $n$ as (a) successively offset by $0.5\,$GHz.
Spectra were taken at $\nu$ intervals of 0.005.
We denote the lower and higher $\nu$ minima respectively by \na\ and \nb.
\textbf{c}, $\sf$ vs $n$, where $S$ is the integrated \resxx\ vs $f$ for the resonance, for $\nu=0.4$ (where the integration is performed for $0.1<f<1.5$\,GHz).
Conversion to $\gamma$ is plotted on the top axis.
The error in $\sf$ is calculated as the difference between the measured value and that obtained by fitting the resonance to a Gaussian.
The dotted line marks the theoretical value \citep{fukuyama:1978} $\sf=\pi e^{2}\nu/2h$ for full charge carrier participation.
Except near $n=1.38$, at which $S/\fpk$ is suppressed, the curve increases before saturating at $\sim 85\%$ of the theoretical value.
The observed increase of $\sf$ demonstrates increase in the strength of the bilayer solid with increasing $n$ as the available particles contribute to the solid.
\textbf{d}, The peak resonance \spk\ vs $\nu$ at fixed $n=1.30$.
Here the well is imbalanced by asymmetric gating to transfer charge $\Delta n$ between the two layers; see Methods for more details.
}
\vspace{-0.2 in}
\label{spks}
\end{figure}
%%%%%%%%%%%%%%%%%%%%%%%%%

The strong pinning mode shown in Fig.\,\ref{wf} is evidence of a solid.
We interpret the two sharp increases in $\fpk$ as due to transitions between different solid configurations.
Figure \,\ref{iplots} gives an overview of the spectra for the $n$ range we surveyed, as image plots of Re\,$(\sigma_{xx})$ in the $(\nu,f)$-plane.

Figures \,\ref{spks}\,(a) and (b) show  \spk\ (Re\,$(\sigma_{xx})$ at the resonance maximum)   and \fpk\ vs $\nu$, respectively, for many $n$.
We interpret the minima in \spk\ vs $\nu$ as due to the transitions, and denote the lower and higher $\nu$ of the  transitions by \na\ and \nb, respectively. 
At \nb\ the  \fpk\  jump is maximal for $n=1.26$, just above the transition to the insulator from the FQHE.
This jump weakens gradually as $n$ increases and is absent by $n=1.41$.

The transitions  within the insulator at \na\ and \nb\ are most pronounced  for  $n$ near the  transition to insulator from the FQHE.   
Except  near \na\ or \nb, the resonance \spk\ increases with $n$, as   seen in  \rfig{spks}\,(a).   
At the same time \fpk\ changes little with $n$ except near \na\ or \nb.  
Pinning modes approximately obey a sum rule \cite{fukuyama:1978,clidensity,murthyreview}, 
 $\pi\nu e^2/2h =  S/\fpk$, where $S$ is the integrated \resxx\ vs $f$ for the resonance.  
\rFig{spks}\,(c) shows   $S/\fpk$ vs $n$ for $\nu=0.4$.   
Except near  $n=1.38$, at which $S/\fpk$ is suppressed, this $\nu$  is   away from \na\ and \nb.
The curve saturates  as $n$ goes above $1.4$  to a value in reasonable agreement with the sum rule; $n\simeq 1.4$ is approximately the largest density at which we observe the jump in \fpk\ at \nb, and the minima in \spk\ vs $\nu$ are considerably weakened by this $n$.

An important result is that the solid appears to be a bilayer.
Figure \ref{spks}\,(d)  shows \spk\ vs $\nu$ for a fixed overall density of $n=1.30$, but varying the charge transfer, $\Delta n$, between layers. 
As layer imbalance   grows, \spk\ is  reduced, and the transition minima shift only slightly in $\nu$.  
The interpretation is that excess carriers of the majority layer do not participate in the resonance, and reduce its strength by damping.  
The dc transport study in Ref.\,\cite{manoharan:1996} found the resistivity also to be sharply reduced on layer imbalance for a 75 nm WQW in this  low $\nu$ insulator.

\begin{figure}[p]
\includegraphics[width=0.5\textwidth]{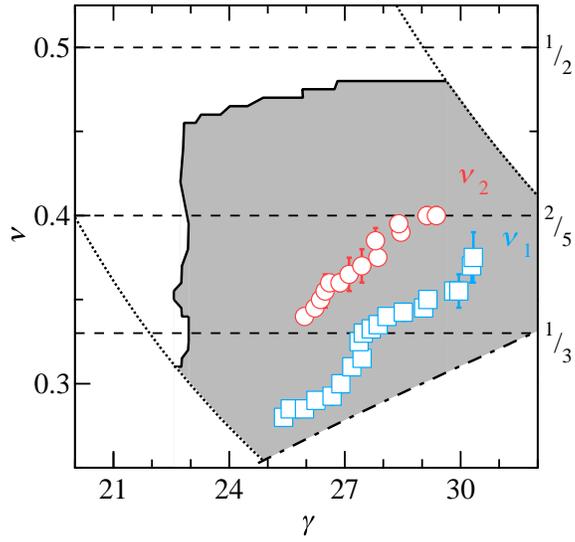}
\vspace{-0.2 in}
\caption{\textbf{Phase diagram in the  $\gamma$-$\nu$ plane.}
The shaded region is where microwave spectra show a resonance of amplitude $\spk> 2\,\mu$S.
The edges of the measurement range resulting from limits in  the  sample density range obtainable from gating  are shown as dotted lines. 
The measurement range limit due to the maximum magnetic field is the upward sloping dashed-dotted line.
The squares and circles denote the  transition filling factors, $\na$ and $\nb$ respectively, as identified from minima in the resonance peak conductivity, \spk,  vs $\nu$.  
}
\label{phase}
\end{figure}

 \rFig{phase}  shows the $\gamma$-$\nu$  plane  with the shaded area marking the region in which a resonance is observed.  
Compared  to the  dc-transport boundary of \rfig{fig1}\,(a) \citep{manoharan:1996},  
the nearly vertical left boundary of the solid in \rfig{phase} occurs at larger  $\gamma$.  
Some of this difference is likely due to the WQW in this paper being wider, 80 nm vs 75 nm for Ref.\,\citep{manoharan:1996}.   
It may be that relative to the dc boundary, the resonance appears  further into the insulating phase.  
This  difference between the resonance boundary and the dc boundary is what reduces the apparent depth of the rightward boundary indentation centered at $\nu=1/3$. We also  find no  resonance in the $\nu>1/2$ reentrant insulator adjacent to the 1/2 FQHE.

 %MS: you need to say a few words about the meaning/significance of d/lb
As yet there is no theory of 2C solids in WQWs.   
Hartree-Fock \cite{narasimhan:1995,fertigbi} and classical \cite{peeters} theories for double quantum wells (DQWs)  admit phase transitions between different types of solids.   
The Hartree-Fock DQW theories \cite{narasimhan:1995,fertigbi} predict multiple phases of bilayer solid, as sketched in \rfig{fig1}\,(b).
A 1C triangular solid exists at low $n$, large interlayer separation $d$ or small $\gamma$, while a weakly-coupled, staggered triangular 2C solid exists at the opposite extreme.
2C staggered square, rhombic or rectangular lattices exist in between, when \dsas\ is small enough and  $d/l_B$, which measures the ratio of intralayer and interlayer interaction energies, is in a particular ($\nu$-dependent) range.     The theories \cite{narasimhan:1995,fertigbi} predict  that $\nu$ of all transitions decreases with increasing $\gamma$, in contrast to the rising \na\ and \nb\ vs $\gamma$ in \rfig{phase}.   Nonetheless the predicted $\nu,\gamma$ of a  solid-solid transition comes close to that of the observed transitions.   For example, for $\nu\sim 1/3$ our observed $\na$ occurs for $\gamma\sim 27.5$.
The lowest-$\gamma$  transition predicted \cite{narasimhan:1995}  between 2C lattices (staggered square to staggered rhombic) is at $\gamma\sim33$, for $d/l_B=3$ and $\nu=1/3$.   
(Ref.\,\citep{narasimhan:1995} calculates up to $d/l_B=3$  and we extrapolate  a  larger $\gamma$ for  $d/l_B\sim 8$, which applies to the transition we see at  $\na=1/3$.)     One source of differences between the theory and the results may be the different vertical confinement 
in WQWs and the DQWs for which the theory was done. 

Theories \cite{archer:2013,archer:2013b} of composite fermion \cite{jainbook} (CF) Wigner crystals  have  considered single-layer systems, but predict transitions between solids of different CF vortex number $2p$.   
Such transitions may explain the evolution \cite{chen:2004} of the pinning  mode for $\nu<1/5$ in lower $n$ single-layer quantum well states, and may be related to the transitions of Ref.\,\citep{hatke:2014}. 
$2p$ is predicted to increase as $\nu$ decreases, where a transition from $2p=2$ to $4$ should occur between $1/5$ and $1/6$.   
This is in the per-layer-$\nu$ range  of the transitions we see, but for the large $\gamma$ case, which corresponds to two weakly coupled parallel layers, the transitions disappear. 
This is contrary to the expectation for the single-layer-like CF vortex-number transitions, which    calculations \cite{archer:2013} indicate are driven by $\nu$, and whose phase diagram in a single-layer quantum well was described  \cite{archer:2013} as not sensitive to well width.

The strengthening of the resonance as $n$  moves deeper into the insulator from the FQHE liquids suggests that the solid may be existing along with a component that does not produce a resonance, and that the solid fraction increases with $\gamma$.  
A series   of intermediate phases with FQHE liquid and Wigner solid components was  proposed theoretically \cite{kivspiv}.   
The closely paired transitions at $\na$ and \nb\ may be between different intermediate phases.   
The disappearance of the transitions at approximately the $n$   at which $S/\fpk$ vs $n$ saturates is also qualitatively suggestive of  the intermediate phase picture, in which the system at large $\gamma$ becomes a homogenous solid.

%\section{Discussion}
%%%%%%%%%%%%%%%%%%%%%%%%%
%fig 5
%%%%%%%%%%%%%%%%%%%%%%%%%

In summary,  microwave  measurements of a WQW  establish its  low $\nu$ insulator  to be  a bilayer solid, since it has a pinning mode resonance that is degraded by layer density imbalance.    
The resonance reveals a pair of structural transitions within the insulator.   
Increases in resonance strength with $\gamma$  and  the disappearance of the transitions as $S/\fpk$ reaches full strength, are qualitatively consistent with a picture of an inhomogeneous insulator containing resonant and nonresonant components.

\section{Methods}
\subsection{Charge Distribution}
The density of the well was changed by front and back gates.
The back gate was in direct contact with the bottom of the sample and the front gate was deposited on a piece of glass that was etched to be spaced from the sample surface to not interfere with the microwave transmission line.
A symmetric, balanced, growth direction charge distribution was maintained by biasing the gates such that individually they would change the density by the same amount with equal and opposite electric fields.
The asymmetric, imbalanced, distribution was obtained by first biasing one gate to get half the desired charge imbalance and then biasing the other in the opposite manner to get the same charge imbalance while maintaining the same total density with applied electric fields in equal amount and the same direction. $d$ and \dsas\ were calculated from simulations.   The calculated  \dsas\ was used to obtain $\gamma$, and had an uncertainty of about $\pm15$ \% for this WQW.

\subsection{Microwave Spectroscopy}
Our microwave spectroscopy  technique \cite{chen:2004,hatke:2014} uses a coplanar waveguide (CPW) on the surface of a sample.
A NiCr front gate was deposited on glass that was etched to space it from the CPW by $\sim 10\,\mu$m.
A schematic diagram of the microwave measurement technique is shown in Fig.\,\ref{fig1}\,(c) and cutaway view of the sample is shown in Fig.\,\ref{fig1}\,(d).
In the high-frequency, low-loss limit,  diagonal conductivity is approximated by  $\sigma_{xx} (f) = (s/ l \zo) \ln (t/\to)$, where $s=30\ \mu$m is the distance between the center conductor and ground plane, $l=28\,$mm is the length of the CPW, and $\zo=50\,\Omega$ is the characteristic impedance without the 2DES.  
$t$ is the amplitude at the receiver and $\to$ is the normalizing amplitude.  The normalizing amplitude was taken  at $\nu=1/2$; the difference  on using $\nu=1$ rather than $1/2$ as a reference is less than $1\,\mu$S.    
We calculate the reported \resxx\ data from $t$ using the results of modeling the fields and currents of the system of the CPW coupled to the WQW, without requiring a high-$f$ low-loss limit.   
The field modeling has been treated elsewhere \citep{fogler:2000,zhwimbal}; it results in a correction of at most a factor of two from the approximate formula, and has no effect on measured peak frequencies.
Hence $\sigma_{xx} (f)$ is the difference between the conductivity and that for $\nu=0.5$.  
The microwave measurements were carried out in the low-power limit,  in which the measurement is not sensitive to the excitation power.

\section{Author Contributions}
A.\,T.\,H. conceived and designed the experiment, performed the microwave measurements, analyzed the data, and co-wrote the manuscript.
Y.\,Liu performed computer simulations, discussed data analysis, and co-wrote the manuscript.
L.\,W.\,E. conceived and designed the experiment, discussed data analysis, and co-wrote the manuscript.
M.\,S. conceived the experiment, discussed data analysis, and co-wrote the manuscript.
L.\,N.\,P., K.\,W.\,W. and K.\,W.\,B. were responsible for the growth of the samples.

\section{Acknowledgements}
We thank Jainendra Jain and Kun Yang for enlightening discussions.
We also thank Ju-Hyun Park and Glover Jones for technical assistance.
The microwave spectroscopy work at NHMFL was supported through DOE grant DE-FG02-05-ER46212 at NHMFL/FSU.   
The National High Magnetic Field Laboratory (NHMFL),  is supported by NSF Cooperative Agreement No. DMR-0654118, by the State of Florida, and by the DOE.   
The work  at Princeton was funded through the NSF through MRSEC DMR-0819860, the Keck Foundation, and the Gordon and Betty Moore Foundation (grant GBMF4420).

\end{document}